\documentclass[traditabstract]{aa}

\usepackage{natbib}
\usepackage{graphicx}
\usepackage{lscape}
\usepackage{times}

\makeatother

\newcommand\av{$A_V$}
\newcommand\nh{$N_{\rm{HI}}$} 
\newcommand\nhx{$N_{\rm{H,X}}$} 
\newcommand\swift{{\it Swift}} 
\newcommand\invsqrcm{cm$^{-2}$} 
\newcommand\Nzn{$N_{\rm ZnII}$} 
\newcommand\Nfe{$N_{\rm FeII}$} 
\newcommand\Nsi{$N_{\rm SiII}$} 
\newcommand\Ns{$N_{\rm SII}$} 
\newcommand\Nnzn{$N_{\rm{N,Zn}}$} 
\newcommand\Nnfe{$N_{\rm{N,Fe}}$} 
\newcommand\Nnsi{$N_{\rm{N,Si}}$} 
\newcommand\Nns{$N_{\rm{N,S}}$} 
\newcommand\Nno{$N_{\rm{N,O}}$} 
\newcommand\Nm{$N_{\rm ntr}$} 
\newcommand\Nox{$N_{\rm O,X}$} 
\def\lesssim{\mathrel{\hbox{\rlap{\hbox{\lower4pt\hbox{$\sim$}}}\hbox{$<$}}}}
\def\gtrsim{\mathrel{\hbox{\rlap{\hbox{\lower4pt\hbox{$\sim$}}}\hbox{$>$}}}}

\begin{document}

\title{The missing gas problem in GRB host galaxies: evidence for a highly ionised component}

\author{P.~Schady\inst{1}, S.~Savaglio\inst{1}, T.~Kr{\"u}hler\inst{1,2}, J.~Greiner\inst{1} \and A.~Rau\inst{1}}
\institute{Max-Planck Institut f{\" u}r Extraterrestrische Physik, Giessenbachstrasse, 85748, Garching, Germany\\\email{pschady@mpe.mpg.de}
\and
Universe Cluster, Technische Universit\"{a}t M\"{u}nchen, Boltzmannstrasse 2, 85748, Garching, Germany
}

\date{Received August 27; accepted}

\abstract
{There is considerable discrepancy between the amount of X-ray absorption and that inferred from optical (rest frame UV) as measured along gamma-ray burst (GRB) sight lines, with the former being typically an order of magnitude higher than what would be expected from the measurement of neutral element species via optical absorption line spectroscopy. We explore this ``missing gas problem" by using X-ray and optical measurements in a sample of 29 z $=0.7-6.3$ GRBs from both spectroscopic data and the afterglow broadband spectral energy distributions. The low ionisation species detected in the UV are associated with the neutral interstellar medium in the GRB host galaxy, while soft X-ray absorption, which is weakly dependent on the ionisation state of the gas, provides a probe of the total column of gas along the sight line. After careful consideration of any systematic effects, we find that the neutral gas consists of $\lesssim 10$\% of the total gas, and this limit decreases with the more ionised that the X-ray absorbing gas is, which in our spectral fits is assumed to be neutral. Only a very small fraction of this ionised gas, however, is detected in UV absorption lines with ionisation potentials up to $\sim 200$~eV (i.e. Si~{\sc iv}, C~{\sc iv}, N~{\sc v}, O~{\sc vi}), which leaves us to postulate that the X-ray excess is due to ultra-highly-ionised, dense gas in the GRB vicinity. 

}

\titlerunning{Properties of the ISM of GRB Host Galaxies}
\authorrunning{Schady et al.}
\maketitle

\section{Introduction}
\label{sec:intro}
The vast luminosities emitted by quasi stellar objects (QSOs) have long been recognised as providing a powerful  tool with which to study the environmental properties of high-redshift galaxies, and ever since the discovery of gamma-ray burst (GRB) afterglows \citep[GRB~970228;][]{vgg+97} GRBs have also been increasingly used to probe distant, otherwise inaccessible environments through their highly luminous emission \citep[e.g.][]{kka+06,bpc+06,fsl+06,rst+07,gkf+09}. Recent years have seen an increase in the abundance of early-time, broadband photometric and spectroscopic data of GRBs, largely owing to the success of the rapid-response, multi-wavelength GRB mission, \swift\ \citep{gcg+04}, and this has revolutionised our capabilities to study in detail the GRB afterglow and the properties of its host galaxy. GRB observations are now reaching further distances than QSOs \citep[e.g. GRB~090423;][]{sdc+09,tfl+09}, and their transient nature allows the host galaxy to be studied at later epochs free of the blinding GRB afterglow \citep[e.g.][]{fls+06,sgl+09}.

There is now a well established connection between long duration GRBs \citep[$\gtrsim 2$~s;][]{kmf+93}, which are the focus of this work, and the formation of massive stars (e.g. GRB980425: Galama et al. 1999, GRB~030329: Hjorth et al. 2003; Stanek et al. 2003). Due to the short life times of massive stars, GRB afterglow observations therefore probe the gas and dust within areas of star formation, in addition to the interstellar medium (ISM) of young, distant galaxies, as well as the intergalactic matter (IGM) intervening these lines of sight.

From the increasing sample of GRB spectroscopic observations that cover the rest frame Ly$\alpha$ feature at 1215~\AA, it is becoming clear that GRB host galaxies have large column densities of cold neutral gas (T$\lesssim 10^3$~K), with the large fraction of GRB hosts containing a damped-Ly$\alpha$ (DLA) system ($\log~N_{\rm HI} > 20.3$~\invsqrcm) or sub-DLA ($19.0 < \log~N_{\rm HI} < 20.3$~\invsqrcm). The survival of certain species, such as Mg~{\sc i}, and time varying Fe~{\sc ii} and Ni~{\sc ii} fine-structure lines place this neutral gas component at a few hundred parsecs from the GRB \citep[e.g.][]{pcb06,vls+07,lvs+09}, within the ISM of the host galaxy.

Absorption by molecular hydrogen in the UV rest frame may provide a probe to the star-forming region closer in to the GRB, where H$_2$ would be expected to exist \citep{tpc+07}. Nevertheless, despite exerted efforts \citep[e.g.][]{vel+04,tpc+07}, molecular hydrogen absorption has, as of yet, only been clearly detected in the spectrum of GRB~080607 \citep{psp+09}\footnotemark[1], and even then the molecular hydrogen may have originated from another molecular cloud along the line-of-sight distinct from the star-forming region surrounding the GRB. However, the number of GRBs with adequate spectral resolution and coverage to detect the H$_2$ Lyman-Werner absorption lines continues to be small, and H$_2$ absorption lines stemming from the GRB star-forming region may still be detected \citep{tpc+07}.
\footnotetext[1]{\citet{fsl+06} reported a tentative detection of H$_2$ in the spectrum of GRB~060206. However, \citet{tpc+07} later argued the claimed H$_2$ absorption lines to be Ly$\alpha$ forest lines.}

For high-z GRBs ($z\gtrsim 2.0$), the circumburst environment may be further probed by the detection of absorption lines produced by highly-ionised species (i.e. ionisation potentials between 50--150~eV) present in the hot gas (T$\sim 10^4$~K) surrounding the GRB. Although these highly ionised absorption lines may also originate from hot gas in the halo of the galaxy, measurements of O~{\sc vi}, C~{\sc iv}, Si~{\sc iv} and N~{\sc v} in a handful of GRB afterglow spectra have been used to probe the gas close in to the GRB progenitor star \citep{flv+08,pdr+08}

In contrast to the specific regions of gas that can be identified from UV spectra, X-ray spectroscopic observations provide measurements of the total column density of gas along the line-of-sight, probing both the cold neutral gas as well as the warm ionised regions. Soft X-rays with energies $< 0.8$~keV are absorbed by medium-weight metals along the line-of-sight, which is predominantly in the form of oxygen and to a lesser extent carbon and nitrogen. Furthermore, the cross-section of oxygen remains relatively unchanged regardless of ionisation state \citep{vy95}, making it a good proxy for the host galaxy total oxygen column density. 

In addition to the conditions of the host galaxy gas along the line-of-sight, the host dust abundance and extinction properties can be explored from dust depletion patterns inferred from relative metal abundances \citep[e.g.][]{sf04}, and also from the analysis of broadband infrared (IR) through to UV spectral energy distributions (SED) \citep[e.g.][]{gw01,sfa+04,kkz06,sww+07, smp+07, spo+10}. Soft X-rays are also extinguished and scattered by dust, although by much smaller amounts such that absorption by medium weight metals is the dominant source of attenuation \citep[e.g.][]{wd01,dra03}.

Due to the complex structure of the absorbing gas and the uncertainties in the location of the various attenuating gas and dust components, a combined analysis of the UV and the soft X-ray energy bands gives a more comprehensive understanding of the properties of the circum- and interstellar medium of GRB host galaxies. For example, the order of magnitude larger equivalent neutral hydrogen column densities typically derived from soft X-ray absorption\footnotemark[2], \nhx, in comparison to the host galaxy neutral hydrogen column densities measured from damped Ly-$\alpha$ absorption, \nh, led \citet{whf+07} to conclude that the former probed a significant column of ionised gas that is transparent to UV photons.
\footnotetext[2]{Assuming solar abundances from \citet{ag89}.}

In this paper we combine the results from the analysis of optical and X-ray spectra, and also of broadband afterglow SEDs, with the aim of providing greater insight on the ionisation state and dust properties of the host galaxy medium along the line-of-sight to GRBs. Instead of neutral hydrogen, we use absorption lines of singly-ionised metals to trace the neutral gas within GRB host galaxies that can then be compared directly with the soft X-ray absorption, which measures primarily the oxygen column density (neutral and ionised), \Nox. This removes the need for hydrogen column density and host galaxy metallicity measurements, which are frequently uncertain or unavailable, thus increasing our sample size. The results presented here contribute towards not only a better understanding on the nature of galaxies hosting GRBs, but also on the environmental conditions within high-redshift galaxies as a whole.

In section~\ref{sec:sample} we provide our sample selection criteria, and we describe our data reduction and analysis in section~\ref{sec:red}. We present our results and discuss their implications in section~\ref{sec:reslts}, and our conclusions are summarised in section~\ref{sec:conc}. All quoted errors throughout the paper are 1-$\sigma$. Throughout this paper we use the cosmic abundances from \citet{ags+09}.

\section{Sample}
\label{sec:sample}
The sample selection is based on the requirement that optical spectroscopic data are available for the GRB afterglow with absorption measurements in at least Zn~{\sc ii}, Si~{\sc ii}, S~{\sc ii} or Fe~{\sc ii} with which to probe the host galaxy neutral gas. All these low ionisation species have ionisation potentials (IPs) slightly larger than hydrogen, and all are frequently measured in GRB optical spectra. Our initial sample amounted 7 pre-\swift\ and 22 \swift\ GRBs, of which 26 had X-ray afterglow spectroscopic observations (all \swift\ GRBs plus four pre-\swift\ GRBs), and 23 had optical/X-ray SED-fit host galaxy visual extinction measurements, \av, (all pre-\swift\ and 16 \swift\ GRBs) available either from the literature, or with sufficient UV through to infrared (IR) data for an SED analysis of our own to be possible. Due to the degeneracy that exists between spectral index and extinction, we only consider those \av\ values derived from broadband optical to X-ray SED modelling. As a consequence of our selection criteria, whereby at least one metal line is detected in the optical spectra, a spectroscopic redshift measurement is available for all GRBs in our sample. 

Our requirement for an optical afterglow spectrum introduces a selection effect against dim or highly extinguished GRBs, either from dust in the host galaxy, Galactic dust, or dust within intervening systems. GRB host galaxy ISM properties are independent of the Galactic or foreground dust extinction, and selection effects biased against GRBs with large foreground extinction should, therefore, not affect our overall results. Similarly, selection effects against redshift ranges that have few prominent absorption lines in the observer frame optical bandpass (the `redshift desert'; $z\sim 1-1.5$) are likely independent of the GRB host environmental conditions, and again should not affect our overall findings.

The dominant selection effect that we need worry about is the bias against GRBs highly extinguished within their host galaxy. Recent studies suggest that between 20--30\% of GRBs are undetected in the optical wavelength range as a result of dust extinction \citep[e.g.][]{ckh+09,fjp+09,pcb+09,gkk+10}, and there is some indication that these more extinguished GRBs reside in more massive and evolved galaxies than the hosts to optically bright GRBs. It should, therefore, be kept in mind that the results presented in this paper apply to GRBs with low host galaxy extinction along the line-of-sight ($A_V\lesssim 1.0$). In the case of selection effects against dim GRBs, our results will be affected if low luminosity GRBs belong to a different class of GRB to their high luminosity counterparts, as suggested by \citet{lz06} and \citet{ngg+06,ngg08} \citep[although see][]{ops+09}, and this possibility should, therefore, be kept in mind.  We, therefore, stress that the results presented in this paper may not apply to low luminosity and/or highly extinguished GRBs.

\section{Data analysis and reduction}
\label{sec:red}
All optical spectroscopic column density measurements used in this paper are listed in Table~\ref{tab:sample} and have mostly ($\sim 75$\%) been taken from the literature, in which case the corresponding reference is also given. In a few cases we re-computed the column densities ourselves using the published equivalent widths (EW) and applying standard techniques. For the 26 GRBs with X-ray spectroscopic data, we preferentially used the X-ray column densities measured from optical to X-ray SED analysis for consistency with the \av\ vaue that we used. Where this was not possible due to the lack of optical/NIR data (five GRBs), we used
\onecolumn
\begin{tiny}
\begin{landscape}
\begin{table*}
\caption{GRB host galaxy column densities for 29 GRBs. Columns list redshift, Galactic column density and visual extinction in the line-of-sight to the GRB, the H~{\sc i}, Si~{\sc ii}, S~{\sc ii}, Fe~{\sc ii} and Zn~{\sc ii} column density with corresponding references, the oxygen column density measured from the X-ray spectra, \Nox, and the best-fit visual extinction, \av, with both with corresponding references given. For GRB~030226 and GRB~090323, two absorbing components were detected within the GBR host galaxy, and the optical column densities for both are given where applicable.\label{tab:sample}}
\begin{center}
\begin{tabular}{lc|cc|lllllll}
\hline\hline
 & & \multicolumn{2}{c}{Galactic parameters} & \multicolumn{7}{c}{GRB host interstellar and circumstellar parameters} \\
GRB & z & $\log N_{\rm H}$ & $A_V$ & $\log N_{HI}$ & $\log $\Nsi & $\log $\Ns & $\log $\Nfe & $\log $\Nzn & $\log $\Nox & \av \\
 & & $\log {\rm (cm}^{-2}{\rm )}$ & mag & $\log {\rm (cm}^{-2}{\rm )}$ & $\log {\rm (cm}^{-2}{\rm )}$ & $\log {\rm (cm}^{-2}{\rm )}$ & $\log {\rm (cm}^{-2}{\rm )}$ & $\log {\rm (cm}^{-2}{\rm )}$ & $\log {\rm (cm}^{-2}{\rm )}$ & mag \\
\hline\hline
990123 & 1.600 & 20.26 & 0.05 & \dots & \dots & \dots & ${14.78^{+0.17 }_{-0.10}}^1$ & ${13.95\pm 0.05}^1$ & ${18.46^{+0.27}_{-0.43}}^2$ & $< 0.25^2$ \\
000926 & 2.038 & 20.35 & 0.07 & ${21.30\pm 0.20}^3$ & ${16.47^{+0.10}_{-0.15}}^1$ & \dots & ${15.60^{+0.20}_{-0.15}}^1$ & ${13.82\pm 0.05}^1$ & $0.0^2$ & ${0.38\pm 0.05}^2$ \\
010222 & 1.475 & 20.22 & 0.07 & \dots & ${16.09\pm 0.05}^1$ & \dots & ${15.32^{+0.15}_{-0.10}}^1$ & ${13.78\pm 0.07}^1$ & ${18.75\pm 0.16}^2$ & ${0.24^{+0.08}_{-0.09}}^2$ \\
020405 & 0.691 & 20.62 & 0.17 & \dots & \dots & \dots & ${15.05\pm 0.75}^4$ & \dots & $< 18.69^5$ & ${0.42\pm 0.02}^5$ \\
020813 & 1.255 & 20.79 & 0.34 & \dots & ${16.29\pm 0.04}^6$ & \dots & ${15.48\pm 0.04}^6$ & ${13.54\pm 0.06}^6$ & \dots & \dots \\
030226$^*$ & 1.987 & 20.20 & 0.06 & ${20.50\pm 0.30}^7$ & ${15.07\pm 0.04}^7$ & $> 13.3^7$ & ${14.860^{+0.002}_{-0.004}}^7$ & $< 12.70^7$ & $< 18.60^5$ & ${0.05\pm 0.01}^5$ \\
 & 1.963 & & & ${20.50\pm 0.30}^7$ & ${14.97\pm 0.03}^7$ & \dots & ${14.479^{+0.002}_{-0.004}}^7$ & ${< 12.60}^7$ & & \\
030323 & 3.371 & 20.61 & 0.15 & ${21.90\pm 0.07}^8$ & ${16.15^{+0.26}_{-0.19}}^8$ & ${15.84\pm 0.19}^8$ & ${15.93\pm 0.08}^8$ & ${< 14.50}^8$ & \dots & \dots \\
050401 & 2.899 & 20.64 & 0.20 & ${22.60\pm 0.30}^9$ & ${16.50\pm 0.40}^9$ & \dots & ${16.00\pm 0.20}^9$ & ${14.30\pm 0.30}^9$ & ${18.47^{+0.18}_{-0.25}}^5$ & ${0.45\pm 0.03}^5$ \\
050505 & 4.275 & 20.23 & 0.06 & ${22.05\pm 0.10}^{11}$ & ${15.70\pm 0.00}^{11}$ & ${16.1}^{11}$ & ${15.50\pm 0.00}^{11}$ & \dots & ${18.93\pm 0.05}^{10}$ & \dots \\
050730 & 3.969 & 20.48 & 0.16 & ${22.10\pm 0.10}^{12}$ & ${15.47\pm 0.03}^{12}$ & ${15.11\pm 0.04}^{12}$ & ${15.31^{+0.04}_{-0.05}}^{12}$ & \dots & ${18.86^{+0.06}_{-0.07}}^{13}$ & $< 0.17^{13}$ \\
050820A & 2.615 & 20.64 & 0.14 & ${21.05\pm 0.10}^{12}$ & ${> 15.43}^{14}$ & ${15.57\pm 0.04}^{14}$ & ${14.82\pm 0.12}^{14}$ & ${12.96\pm 0.02}^{14}$ & ${18.41^{+0.07}_{-0.05}}^{13}$ & ${0.27\pm 0.04}^{13}$ \\
050904 & 6.295 & 20.66 & 0.18 & $21.62\pm 0.20^{15}$ & $< 16.60^{16}$ & ${15.95\pm 0.55}^{16}$ & \dots & \dots & ${19.11^{+0.09}_{-0.11}}^{10}$ & \dots \\
050922C & 2.199 & 20.73 & 0.32 & ${21.55\pm 0.10}^{12}$ & ${14.97\pm 0.07}^{17}$ & ${14.87\pm 0.05}^{12}$ & ${14.58^{+0.26}_{-0.17}}^{17}$ & \dots & ${17.92^{+0.20}_{-0.33}}^{13}$ & ${0.14^{+0.08}_{-0.07}}^{13}$ \\
051111 & 1.549 & 20.72 & 0.50 & \dots & ${> 16.14}^{14}$ & \dots & ${15.32\pm 0.01}^{14}$ & ${13.47\pm 0.04}^{14}$ & ${18.47^{+0.07}_{-0.13}}^{13}$ & ${0.39^{+0.11}_{-0.10}}^{13}$ \\
060206 & 4.048 & 19.95 & 0.04 & ${20.85\pm 0.10}^{18}$ & ${15.23\pm 0.04}^{19}$ & ${15.13\pm 0.05}^{19}$ & ${> 14.65}^{20}$ & \dots & ${18.82^{+0.14}_{-0.13}}^{13}$ & $< 0.24^{13}$ \\
060418 & 1.490 & 20.94 & 0.69 & \dots & ${15.92\pm 0.03}^{21}$ & \dots & ${15.22\pm 0.03}^{14}$ & ${13.09\pm 0.01}^{21}$ & ${18.32^{+0.16}_{-0.17}}^{13}$ & ${0.13^{+0.01}_{-0.02}}^{13}$ \\
060510B & 4.941 & 20.61 & 0.12 & ${21.30\pm 0.10}^{22}$ & \dots & ${15.60^{+0.30}_{-0.80}}^{23}$ & ${> 15.90}^{23}$ & \dots & ${19.07^{+0.14}_{-0.12}}^{10}$ & \dots \\
060526 & 3.221 & 20.70 & 0.21 & ${20.01\pm 0.15}^{24}$ & ${> 14.48}^{24}$ & ${14.58\pm 0.25}^{24}$ & ${14.28\pm 0.24}^{24}$ & ${< 12.73}^{24}$ & $< 18.12^{13}$ & $< 0.59^{13}$ \\
070802 & 2.450 & 20.46 & 0.08 & $21.50\pm 0.20^{25}$ & ${16.41\pm 0.17}^{25}$ & \dots & ${16.44\pm 0.12}^{25}$ & ${13.97\pm 0.14}^{25}$ & ${18.99^{+0.13}_{-0.10}}^{26}$ & ${1.23^{+0.18}_{-0.16}}^{26}$ \\
071003 & 1.604 & 21.03 & 0.47 & \dots & \dots & \dots & ${13.77^{+0.38}_{-0.17}}^{27}$ & ${< 12.75}^{27}$ & ${17.37^{+0.15}_{-0.39}}^{10}$ & \dots \\
071031 & 2.692 & 20.09 & 0.04 & ${22.15\pm 0.05}^{12}$ & \dots & \dots & ${15.83\pm 0.02}^{12}$ & ${13.02\pm 0.02}^{12}$ & ${18.69^{+0.28}_{-0.30}}^{26}$ & $< 0.11^{26}$\\
080319B & 0.937 & 20.05 & 0.03 & \dots & \dots & \dots & ${14.56^{+0.05}_{-0.04}}^{28}$ & \dots & ${17.74^{+0.14}_{-0.17}}^{13}$ & ${0.06^{+0.03}_{-0.02}}^{13}$ \\
080330 & 1.511 & 20.09 & 0.05 & \dots & ${> 15.41}^{29}$ & \dots & ${14.70\pm 0.09}^{29}$ & ${12.79\pm 0.06}^{29}$ & ${18.18^{+0.20}_{-0.26}}^{26}$ & $< 0.19^{26}$ \\
080413A & 2.433 & 20.94 & 0.48 & ${21.85\pm 0.15}^{12}$ & \dots & \dots & ${15.57\pm 0.04}^{12}$ & ${12.85\pm 0.06}^{12}$ & ${18.64^{+0.21}_{-0.31}}^{13}$ & $< 0.88^{13}$ \\
080607 & 3.036 & 20.23 & 0.07 & ${22.70\pm 0.15}^{30}$ & $> 16.20^{29}$ & $> 16.14^{30}$ & $> 16.35^{30}$ & $> 13.60^{30}$ & ${19.11^{+0.05}_{-0.03}}^{10}$ & \dots \\
080721 & 2.591 & 20.84 & 0.31 & ${21.60\pm 0.10}^{31}$ & $> 16.00^{31}$ & \dots & \dots & \dots & ${18.69^{+0.06}_{-0.07}}^{13}$ & ${0.39^{+0.16}_{-0.14}}^{13}$ \\
090323 & 3.577 & 20.23 & 0.08 & ${19.62\pm 0.33}^5$ & ${15.46\pm 0.13}^5$ & ${15.41\pm 0.04}^5$ & ${14.91\pm 0.05}^5$ & $< 12.7^5$ & $< 18.65^5$ & ${0.10\pm 0.04}^5$ \\
 & 3.567 & & & ${20.72\pm 0.09}^5$ & ${15.80\pm 0.05}^5$ & ${15.80\pm 0.02}^5$ & ${15.00\pm 0.05}^5$ & ${13.57\pm 0.04}^5$ & & \\
090328 & 0.736 & 20.70 & 0.18 & \dots & \dots & \dots & ${16.07^{+0.53}_{-0.36}}^{32}$ & \dots & ${{18.39}^{+0.11}_{-0.12}}^5$ & ${0.34\pm 0.04}^5$ \\
090926A & 2.106 & 20.44 & 0.07 & $21.73\pm 0.07^{33}$ & ${14.97\pm 0.26}^{33}$ & ${14.96\pm 0.15}^{33}$ & ${14.36\pm 0.14}^{33}$ & ${< 13.00}^{33}$ & $< 18.74^5$ & $< 0.1^5$ \\
\hline
\end{tabular}
\end{center}
$^1$ \citet{sff03};
$^2$ \citet{sww+07};
$^3$ \citet{hmg+03}; 
$^4$ \citet{mpp+03};
$^5$ This work;
$^6$ \citet{sf04};
$^7$ \citet{sbp+06};
$^8$ \citet{vel+04};
$^9$ \citet{wfl+06};
$^{10}$ \citet{ebp+09};
$^{11}$ \citet{bpc+06};
$^{12}$ \citet{lvs+09};  
$^{13}$ \citet{sdp+10};
$^{14}$ \citet{pcb+07};
$^{15}$ \citet{tot06};
$^{16}$ \citet{kka+06};
$^{17}$ \citet{pwf+08};
$^{18}$ \citet{fsl+06};
$^{19}$ \citet{twl+08};
$^{20}$ \citet{ath+09};
$^{21}$ \citet{vls+07};
$^{22}$ \citet{cbc07};
$^{23}$ \citet{psc+07};
$^{24}$ \citet{tkj+08};
$^{25}$ \citet{efh+09};
$^{26}$ \citet{gkk+10};
$^{27}$ \citet{plc+08};
$^{28}$ \citet{dfp+09};
$^{29}$ \citet{dfp+09};
$^{30}$ \citet{psp+09};
$^{31}$ \citet{srv+09};
$^{32}$ \citet{mcb+10};
$^{33}$ \citet{rsk+10}\\
$^*$ Where two host galaxy absorption systems are reported, we used the combined column densities of the two systems in our analysis.
\end{table*}
\end{landscape}
\end{tiny}
\twocolumn
X-ray absorption values measured from the X-ray data alone, which we took from \citet{ebp+09}.

Both in our own X-ray analysis and for the results taken from the literature, the X-ray fitting package {\sc xspec} \citep{arn96} was used to model the X-ray spectra with an absorbed power-law ({\it phabs*pow}), where solar abundances were assumed. This provided an equivalent neutral hydrogen column density, \nhx, which we converted back into an oxygen column density, \Nox, and also list in Table~\ref{tab:sample} with corresponding references where relevant\footnotemark[3]. We note that in using the photoelectric absorption model {\it phabs} we assume that the soft X-ray absorbing gas is neutral, which may not be an accurate approximation \citep[e.g.][]{whf+07,pdr+08}, in which case the column of absorbing material would be underestimated. The best-fit \nhx, and consequently also \Nox\ listed in Table~\ref{tab:sample}, are thus effective lower limits. Furthermore, for increasing redshift, the absorption from oxygen is increasingly shifted out of the bandpass such that at $z> 2.5$ the majority of the soft X-ray absorption is caused by other heavier elements. In this case our oxygen column density measurements are indirectly determined assuming solar relative metal abundances, which is an assumption typically made in X-ray spectroscopic analysis. However, the implications of such an assumption apply as much to \nhx\ as they do to \Nox, but in the latter case a lot of the uncertainty corresponding to the metallicity of the environment for the large fraction of GRBs at $z>2.5$ is removed.

In the final column of Table~\ref{tab:sample} we give the best-fit host galaxy visual extinction along the line-of-sight to the GRB, \av, with relevant references provided when taken from the literature. Finally, the GRB redshift, and the Galactic soft X-ray absorption and visual extinction taken from \citet{kbh+05} and \citet{sfd98}, respectively, is also given in Table~\ref{tab:sample}.
\footnotetext[3]{By default {\sc xspec} uses the abundances from \citet{ag89}, which are consistent with those of \citet{ags+09} within errors. Variations in \nhx\ and \Nox\ measurements resulting from the use of different abundances are, therefore, negligible.}

\subsection{SED Analysis}
\label{ssec:sed}
For consistency within our sample we require that the host galaxy visual extinction and soft X-ray absorption are measured in the same way for each GRB. In \citet{spo+10}, \citet{gkk+10} and \citet{sww+07}, host extinction and absorption values were measured from the same model fits to the GRB SEDs, which we describe in the next section. We, therefore, consider the results from these papers comparable to each other. For six GRBs in our sample not analysed in any of the papers listed above, sufficient optical/NIR and X-ray data was available to produce a broadband SED. In these cases we therefore performed our own SED analysis following the method described in \citet{spo+10}. 

GRB~020405 and GRB~030226 had available {\it Chandra} data, which we reduced using the standard procedure. The optical data for these SEDs was taken from the literature (GRB~020405; Masetti et al. 2003, GRB~030226; Klose et al. 2004). The remaining four GRBs (GRB~050401, GRB~090323, GRB~090328, GRB090926A) had \swift\ XRT \citep{bhn+05} and UVOT \citep{rkm+05} data, which were taken from the UK \swift\ data archive\footnotemark[4], as well as {\it g'r'i'z'} and JHK data taken with the GROND, a simultaneous 7-channel imager \citep{gbc+08} mounted at the 2.2m MPI/ESO telescope at La Silla (Chile). All XRT data were in photon counting (PC) mode and were reduced with the {\sc xrtpipeline} tool (v0.12.4). Spectra were extracted following the method described in \citet{ebp+09}, avoiding time intervals that showed evidence of spectral evolution, such as during the early-time steep decay phase of the X-ray light curve \citep{nkg+06}, during flares \citep{fmr+07}. Both UVOT and GROND data were reduced following the standard procedures described in \citet{pbp+08} and \citet{ykg+08} respectively.
\footnotetext[4]{http://www.swift.ac.uk/swift\_portal/}

The SEDs were fitted within {\sc xspec} (v12.6.0) with a Galactic and host absorbed and extinguished power law, allowing for a cooling break between the X-ray and optical energy bands \citep[with change in spectral slope fixed to $\Delta\beta = 0.5$;][]{spn98}  when it provided a significantly improved fit. Neutral hydrogen absorption from the IGM was also taken into account in our model using the prescription from \citet{mad95}. Solar abundances are assumed in the soft X-ray absorption measurements, and the host visual extinction was measured from the best-fit to the data between the Small or Large Magellanic Cloud (SMC and LMC, respectively), or the Milky Way mean extinction laws \citep{pei92}.

\subsection{Optical Spectroscopy Analysis}
For the $\sim 25$\% of the sample where we performed our own optical spectroscopic analysis, column densities of metals were derived by the curve of growth (COG) analysis \citep{spi78}, or if appropriate, the apparent optical depth (AOD)\citep{ss91} method. The COG provides reliable results through the linear approximation when the equivalent width of a line is EW $< 0.1$\AA\ and the effective Doppler parameter is $b > 20$~km~s$^{-1}$. For stronger lines, the COG requires the detection of at least several lines with different oscillator strengths of the same ion. We, therefore, used the AOD method for moderate to low EWs, provided that the effective Doppler parameter is not very low \citep[$b\gtrsim 10$~km~s$^{-1}$;][]{pcb+07}.

\section{Results and Discussion}
\label{sec:reslts}

\subsection{Dust Within the GRB Host Galaxy}
\label{ssec:GRBdst}
Iron is a highly refractory element and its relative abundance with non-refractory, iron-peak elements acts as a good tracer of the abundance of dust particles present within the region of absorbing gas. The column density of singly ionised iron and zinc are often used for this purpose since both are frequently measured in UV spectra and they both trace neutral gas. It has been suggested that a relative zinc-to-iron abundance of up to [Fe/Zn]\footnotemark[5]$\approx -0.3$~dex could arise in low-metallicity environments due to differences in the nucleosynthesis processes from solar \citep{pw02}. However, any zinc-to-iron relative abundances lower than this are likely to be indicative of dust depeletion.
\footnotetext[5]{[X/Y] = $\log({\rm N_X/N_Y}) - \log({\rm X/Y})_{\odot}$}

\begin{figure}
\centering
\includegraphics[width=0.5\textwidth]{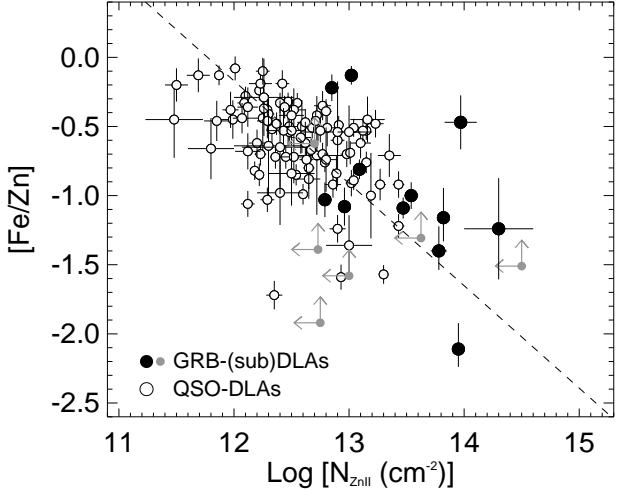}
\caption{Iron-to-zinc relative abundance, [Fe/Zn], as a function of the logarithmic Zn~{\sc ii} column density. GRB host galaxy measurements are plotted as solid circles and a sample of QSOs taken from \citet{sav06} (open circles) are plotted for comparison. The dashed line shows the ordinary least-squares (OLS) bisector fit to the combined GRB and QSO data, which has a power law index of -0.7. GRBs with limits on either Fe~{\sc ii} of Zn~{\sc ii} are plotted in grey and with smaller symbols.
}\label{fig:FeZnvsZn}
\end{figure}

In Fig.~\ref{fig:FeZnvsZn} we plot the iron-to-zinc relative abundance [Fe/Zn] as a function of the Zn~{\sc} column density, \Nzn, for 18 GRBs in our sample with Fe~{\sc ii} and Zn~{\sc ii} column density measurements, where at least one is well constrained (i.e. not an upper or lower limit). Of these 18 GRBs, five have only an upper limit on the Zn~{\sc ii} column density and these are plotted with smaller symbols, and also with left- and upward pointing arrows. A sample of QSOs taken from \citet{sav06} are also plotted with open circles for comparison, and the dashed line is the line of best-fit to the combined GRB and QSO sample. Given the increasing evidence indicating that the bulk of neutral gas within GRB host galaxies probed by optical afterglow spectra lies at a few hundred parsecs from the GRB, the similarity in trend of smaller iron-to-zinc relative abundances for larger Zn~{\sc ii} column densities between QSOs and GRBs is not so surprising. A Spearman rank test for the 13 GRBs with well measured Fe~{\sc ii} and Zn~{\sc ii} column density measurements gives a Spearman coefficient of $\rho=-0.51$ at 90\% confidence, and the Spearman rank coefficient remains the same when we include the QSO data, although the confidence in the coefficient increases, indicating a significant anti-correlation between [Fe/Zn] and \Nzn, and thus suggesting  that dust depletion tends to be higher for larger Zn~{\sc ii} column densities. 

For all 13 GRBs that have well-constrained iron and zinc column densities, seven also have measurements of the host galaxy visual extinction, \av, at the $3\sigma$ level (see Table~\ref{tab:sample}). To test whether the correlation between dust depletion and Zn~{\sc ii} column density indicated by Fig.~\ref{fig:FeZnvsZn} is in agreement with the best-fit \av\ values from our SED analysis, we split this subsample of seven GRBs into two, depending on their Zn~{\sc ii} column density. We use a threshold of \Nzn = 13.5, which is the mean Zn~{\sc ii} column density for the GRB subsample, and find that for the set with $\log~N_{\rm ZnII}< 13.5$ and $\log~N_{\rm ZnII} > 13.5$, the mean best-fit \av\ values are $\langle A_V\rangle = 0.17$ and $\langle A_V\rangle = 0.62$, respectively, with respective standard deviations of 0.07 and 0.27. Although the scatter in \av\ is large, it is reassuring that the trend of larger dust-depletion for larger Zn~{\sc ii} column densities observed in our optical spectroscopic analysis is similarly present in our SED analysis results.

\begin{figure}
\centering
\includegraphics[width=0.5\textwidth]{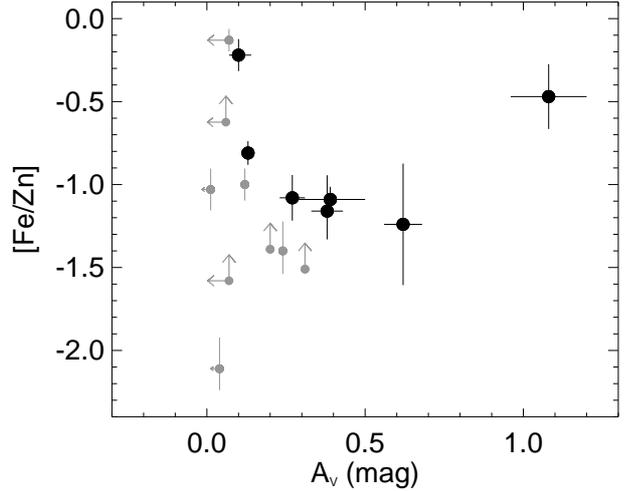}
\caption{Iron-to-zinc relative abundance, [Fe/Zn], against visual extinction, \av, for a sample of 16 GRBs. GRBs with limits on either [Fe/Zn] or \av\ are plotted in grey and with smaller symbols.}\label{fig:AvvsFeZn}
\end{figure}

This general agreement is better illustrated in Fig.~\ref{fig:AvvsFeZn}, where we have plotted the relative iron-to-zinc abundance, [Fe/Zn], against visual extinction, \av, for 16 of the GRBs included in Fig.~\ref{fig:FeZnvsZn} that have \av\ measurements available. GRBs with lower limits in either [Fe/Zn] or \av\ are plotted with smaller symbols, and GRBs with SEDs best-fit with a spectral break between the optical and X-ray energy bands are plotted as open circles. There is one clear outlier in this GRB sample with a large visual extinction of $A_V > 1.0$ but small [Fe/Zn]. This data point corresponds to GRB~070802, and an abundance of dust along its line-of-sight, as suggested by it's large afterglow extinction, may thus no longer make our assumption of negligible zinc depletion valid. Further evidence of the line-of-sight to this GRB having significantly amounts of dust is in the clear detection of a 2175\AA\ dust extinction feature in its optical spectrum \citep{kkg+08,efh+09}\footnotemark[6], 
\footnotetext[6]{GRB~080607 also had a well-detected 2175~\AA\ feature \citep{psp+09}. However, only lower limits on the zinc and iron column densities are available for this GRB.}
which is thought to be more prominent in dusty environments \citep[e.g.][]{npp+07,psp+09}. Although along numerous lines-of-sight the fraction of zinc in grains is generally negligible, it can be as high as 20\% in the cool disc ISM \citep{ss96}, where the dust abundance is higher. The inconsistencies between [Fe/Zn] and \av\ for GRB~070802 may therefore be due to a significant amount of both Fe and Zn dust depletion, causing the relative iron-to-zinc metal abundance to only be a little lower than solar abundances.

\begin{figure}
\centering
\includegraphics[width=0.5\textwidth]{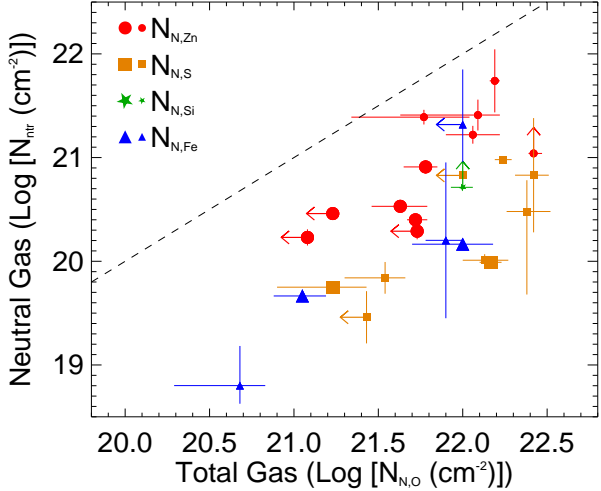}
\caption{Host galaxy neutral gas column density, \Nm, against host galaxy total gas column density, \Nno, along the line-of-sight to a sample of 26 GRBs. For each GRB, \Nm\ is derived from either the column density of Zn~{\sc ii} (red circles), S~{\sc ii} (orange squares), Si~{\sc ii} (green stars) or Fe~{\sc ii} (blue triangles), where a correction for dust depletion is applied to \Nnsi\ and \Nnfe. Smaller data points correspond to those taken from low- or mid-resolution spectra, and larger data points are taken from high-resolution spectra (R $> 10,000$). The dashed line corresponds to where \Nm\ is equal to \Nno.}\label{fig:MNox}
\end{figure}

\subsection{Gas Ionisation State}
\label{ssec:feo}
\subsubsection{Total vs. Neutral Gas}
\label{sssec:totvsntrl}
To trace the neutral gas within a GRB host galaxy, we preferentially use the ion Zn~{\sc ii} or S~{\sc ii}, which if not available is substituted by Si~{\sc ii} or otherwise Fe~{\sc ii}. This is  due to the inherent uncertainties present in the dust depletion of the refractory elements Si~{\sc ii}, and especially Fe~{\sc ii}. When column densities for both Zn~{\sc ii} and S~{\sc ii} are available, the one with the smaller error bars is used. When only Si~{\sc ii} or Fe~{\sc ii} absorption measurements are available, a dust correction is applied to the measured column density. The general consistency between the dust depletion inferred from iron-to-zinc relative abundances and the SED best-fit visual extinction shown in the previous subsection gives confidence to the assumption that zinc dust depletion is typically negligible. We would expect to see a discrepancy between [Fe/Zn] and \av\ where this assumption no longer applies (see Fig.~\ref{fig:AvvsFeZn}). To correct for the dust depletion of iron and silicon, we therefore fitted a linear function to $\log N_{\rm{SiII}}$ and $\log N_{\rm{FeII}}$ against $\log N_{\rm{ZnII}}$, respectively, for the combined data set of GRBs and QSOs, and used these lines of best-fit to correct for dust depletion in our \Nsi\ and \Nfe\ measurements. We then corrected for cosmic abundance variances by normalising all our measurements to each other using the abundances from \citet{ags+09}. We denote the normalised column densities of oxygen, zinc, silicon, sulphur and iron by $N_{N,O}$, $N_{N,Zn}$, $N_{N,Si}$, $N_{N,S}$ and $N_{N,Fe}$, respectively.

In Fig.~\ref{fig:MNox} we plot the neutral gas column density, \Nm, represented by either \Nnzn\ (red circles), \Nns\ (orange squares), \Nnsi\ (green stars) or \Nnfe\ (blue triangles), against \Nno, which represents the total column of gas (neutral and partially ionised), where both \Nnsi\ and \Nnfe\ have been corrected for dust depletion. We have also differentiated between the resolution of the optical spectra used for each GRB observation, where those measurements taken from high-resolution spectra (R $> 10,000$) are plotted with larger symbols, and measurements from low- or mid-resolution spectra are have smaller symbols. Just under two thirds of our sample have low- or mid-resolution spectral data, and this may have consequences for the precision of the column density measurements in these cases, which we discuss further below. The dashed line in Fig.~\ref{fig:MNox} corresponds to where the column density of neutral gas is equal to the total column density of gas as probed by our X-ray absorption measurements. From this figure it is clear that for all GRBs, \Nno\ is larger than the column density of neutral gas, \Nm, by typically around an order of magnitude.

\subsubsection{Systematic Effects}
\label{sssec:seleff}
The neutral gas column density in two thirds of the GRBs plotted in Fig~\ref{fig:MNox} was measured from Fe~{\sc ii} and Zn~{\sc ii} absorption, and in these cases at least some of the excess absorption measured in X-rays could be due to GRB hosts having $\alpha$-element overabundances. $\alpha$-elements are produced within the sequence of He, such as O, Mg, Si and S, and trace massive star formation. An over-abundance of $\alpha$-elements should, therefore be expected in young, star-forming galaxies such as GRB hosts \citep[e.g.][]{pcd+07}. However, such overabundances would not account for the order of magnitude differences between X-ray and optical absorption measurements, nor would it apply to the third of the sample with neutral gas column densities traced by Si~{\sc ii} absorption. We, therefore, do not consider an abundance in $\alpha$-elements within GRB host galaxies to be the dominant cause of the larger values of \Nno\ measurements compared to \Nm.

Alternatively, saturation and lack of resolution in the optical spectral data could lead to the metal column densities being underestimated \citep[e.g][]{pro06}. However, this is not a problem when high-resolution spectra are available, which is the case for $\sim 40$\% of our sample. Optical column density measurements for this fraction of the sample are plotted with larger symbols in Fig.~\ref{fig:MNox}, and most of these fall well below the dashed line, in the region of space where the total gas column density is significantly larger than the neutral gas column density. For the remaining $\sim 60$\% of the sample, the error on the neutral gas column density measured from low- and mid-resolution optical spectra may be as large as 0.6~dex, although this would be a highly pessimistic case. However, even in this unlikely situation, over half of the mid- and low-resolution data points would still have a total gas column density at least an order of magnitude larger than the neutral gas column density. We therefore conclude that although the resolution of the optical spectroscopic data available may cause the neutral gas column density to be under-predicted, for the majority of the data points in Fig.~\ref{fig:MNox}, the neutral gas column density would remain significantly smaller than the metal column density implied by the X-ray data.

A clear possibility that we need to address is that of unaccounted intervening absorbing systems in our line-of-sight, which would result in the GRB host soft X-ray absorption being over-estimated. In optical spectroscopy the location of any intervening absorption systems are clearly determined, and thus the absorption from such systems is easily separated from absorption within the GRB host galaxy. However, the resolution of our X-ray spectra is insufficient for the contribution from several absorbing systems on the GRB afterglow soft X-ray absorption to be disentangled. It should be said that such strong intervening absorbing systems are rare, and the probability of there being an absorption system in the line-of-sight to a GRB of sufficient optical depth to contribute a measurable amount of soft X-ray absorption is, therefore, small \citep{oj97,crc+06}. In \citet{whf+07} this issue was looked at in greater detail, and they estimated that the absorption resulting from any intervening system would need to be two orders of magnitude larger than the absorption measured thus far in optical spectra; specifically than the absorption measured from Mg~{\sc ii} systems, which are typically the largest of absorbers. Further to this, \citet{whf+07} noted the soft X-ray absorption in the line-of-sight to blazars to be small relative to that of GRBs. This is despite the two types of sources having comparable numbers of intervening Mg~{\sc ii} absorption systems along the line-of-sight, thus further supporting an intrinsic absorption interpretation.

A greater worry in our X-ray measurements is that of the Galactic absorption, which in the case of our GRB sample can be uncertain by up to 15\% \citep{kbh+05}. Any under-estimate of the Galactic soft X-ray absorption would result in the host galaxy soft X-ray absorption being over-estimated by a factor of $(1+z)^{2.6}$ \citep{gw01}. In order to investigate whether such a factor could account for the large values of \Nno\ observed in Fig.~\ref{fig:MNox}, we re-determined the host galaxy X-ray absorption for each GRB plotted in the figure, but this time setting the Galactic absorption along the line-of-sight to its $3\sigma$ upper limit. We found the re-calculated host galaxy X-ray absorption values to have typically decreased by a little less than 50~\%, thus remaining a factor of five or so larger than \Nm. Any uncertainties in our Galactic absorption corrections therefore have no significant impact on our primary result that \Nno\ is a factor of several larger than \Nm.

We therefore conclude that the offset between \Nno\ and \Nm\ within GRB host galaxies is predominantly due to X-ray absorption measurements probing a larger column of gas than the UV low ion absorption line measurements.  This larger column of gas must be ionised, which is not traced by the low ionisation species that we have analysed, but which does absorb photons in the soft X-ray energy bands. Dust extinction of soft X-rays may also contribute to the large X-ray inferred column densities. However, due to the effect of self-shielding, the fraction of soft X-ray photons lost to extinction will be significantly below the fraction absorbed by oxygen, carbon and nitrogen \citep[e.g.][]{dra03}. Furthermore, given that the soft X-ray absorption column densities were measured from a neutral absorption model fit to the X-ray spectra, a large ionised gas absorption component would turn the X-ray column density measurements into lower limits, as already indicated in section~\ref{sec:red}. The order of magnitude difference between the column density of neutral gas and the total gas column density measured from the X-ray data would thus suggest that, at least along the line-of-sight to the GRB, $\sim 90$~\% of the gas within GRB host galaxies is ionised.

In at least some GRBs most of the neutral gas lies at a few hundred parsecs from the GRB \citep[e.g.][]{pcb06,vls+07}, suggesting that little neutral gas remains within the star forming region in which the GRB is likely to be embedded. It may therefore not be so surprising that the ratio between ionised-to-neutral gas is so high. As an illustrative example, if we assume a GRB to lie within a dense star-forming region of radius 20~pc then a hydrogen density $\rho\approx 50$~cm$^{-3}$ within this region would be needed to account for our soft X-ray observations, which is not unreasonable. Such a scenario provides a simple explanation to the larger column densities of soft X-ray absorbing gas compared to the neutral gas column densities. It is also consistent with the association of long GRBs with massive star formation, whereby GRBs are expected to lie within dense star-forming regions of the host galaxy, as well as being compatible with the neutral gas lying several hundred parsecs from the GRB, as indicated by optical spectroscopy of some GRB afterglows.

In both \citet{whf+07} and \citet{pdr+08} the quoted distances from the GRB out to which gas could be ionised were smaller than the radius of 20~pc that we consider above (3~pc and 10~pc, respectively). However, these smaller distances were based on the assumption that the GRB was the sole source of ionising photons. When considering the contribution from other massive stars in the neighbourhood, the radius of the ionisation front could increase out to $> 30$~pc from the GRB \citep[][although \citet{whf+07} argued there only to be a small increase in size of the ionised region]{pdr+08}. Nevertheless, even in the case where the size of the ionised region is within 3--10~pc of the GRB, a particle density of $\rho\approx 100-250$~cm$^{-3}$, which remains within reason, would similarly satisfy our observations.

\subsubsection{Highly Ionised Gas}
\label{sssec:highionobs}
To investigate this scenario further, we take a look at the column densities of metals in a higher ionisation state than has been looked at so far. Whereas low ion elements will only exist far from the GRB, such as Zn~{\sc ii}, Si~{\sc ii} and Fe~{\sc ii}, more highly ionised metals, such as C~{\sc iv}, Si~{\sc iv}, N~{\sc v} and O~{\sc vi} could survive within the circumburst environment of the GRB \citep{flv+08,pdr+08}, and may thus make up a large fraction of the ionised gas probed by our soft X-ray observations. Both \citet{flv+08} and \citet{pdr+08} used the measurements of highly ionised metals to probe the environmental conditions close to the GRB for a combined sample of nine GRBs, where the former used the column densities of  C~{\sc iv}, Si~{\sc iv}, N~{\sc v} and O~{\sc vi} to trace the highly ionised gas, and \citet{pdr+08} focused on measurements of N~{\sc v}, which are typically less saturated. 

\begin{figure}
\centering
\includegraphics[width=0.5\textwidth]{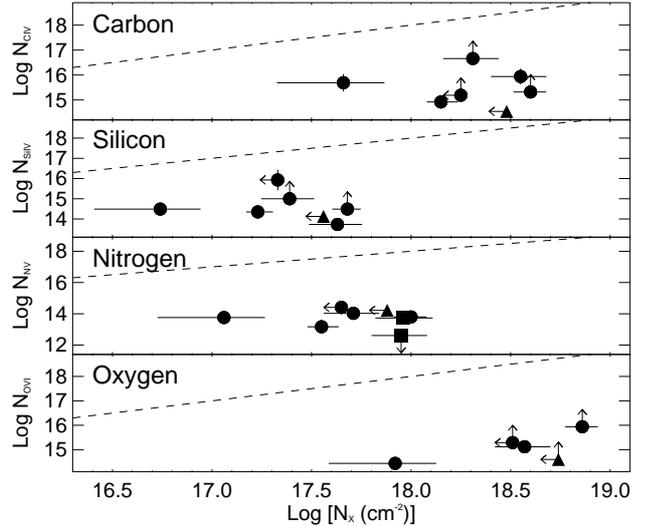}
\caption{Logarithmic host galaxy column density of the highly ionised atoms C~{\sc iv}, Si~{\sc iv}, N~{\sc v} and O~{\sc vi} against the total logarithmic column densities of C, Si, N and O in the top, second, third and bottom panels, respectively. The total column densities are derived from X-ray observations, denoted by $N_X$. The C~{\sc iv}, Si~{\sc iv}, N~{\sc v} and O~{\sc vi} measurements plotted as circles are all taken from \citet{flv+08}, the N~{\sc v} data plotted as squares are from \citet{pdr+08}, and data taken from \citet{dfc+10} are plotted as triangles. In all four panels, the dashed line corresponds to where the normalised soft X-ray column density is equal to the column density of the corresponding highly ionised atom.}\label{fig:highionsvsNmx}
\end{figure}

To quantify the fraction of gas probed by the soft X-ray energy bands that could be in a highly ionised state selected the sample of eight GRBs from \citet{flv+08} and \citet{pdr+08} that had soft X-ray absorption measurements, as well as GRB~090926A, which had high-ion column densities measurements reported \citet{dfc+10}. This sample allows us to compare the soft X-ray column densities to the highly ionised column densities of N, C, Si and O. To do this we normalised the soft X-ray absorption measurements to the cosmic abundance of the element in question being compared (i.e. either Si, C, N or O). Our results are shown in Fig.~\ref{fig:highionsvsNmx}, which compares the logarithmic column density of the highly ionised atoms $N_{CIV}$, $N_{SIV}$, $N_{NV}$ and $N_{OVI}$ against the soft X-ray absorption column densities normalised to C, Si, N and O cosmic abundances in the top, second, third and bottom panel, respectively. The normalised soft X-ray absorption column density is denoted by $N_X$ to indicate the column density of C, Si, N or O as determined from our soft X-ray absorption measurements. The column densities of highly ionised atoms taken from \citet{flv+08} are plotted as circles, the N~{\sc v} measurements from \citet{pdr+08} are shown as squares, and those data from \citet{dfc+10} are plotted as triangles. In all four panels the dashed line represents where the normalised soft X-ray column density measurements are equal to the column density of the highly ionised atom represented in each corresponding panel.

It is clear from this figure that all data points lie several orders of magnitude below the dashed lines, indicating that the highly ionised gas within GRB hosts only makes up a small fraction ($< 0.01$\%) of the soft X-ray absorption. Fig.~\ref{fig:MNox}, on the other hand, indicates that $\sim 90$\% of the gas probed by the soft X-rays is ionised. If the large fraction of this gas is in a lower ionised state, then there should be a signature of this in the detection of strong absorption lines from intermediate ionisation lines with IPs between $\sim 20$~eV and $\sim 50$~eV (i.e. Si~{\sc ii} and Si~{\sc iv}). This is, however, not the case. For example, Al~{\sc iii}, which has an IP of 28~eV, is frequently observed to be weaker than Al~{\sc ii}, which has an IP of 19~eV. This would, therefore, suggest that most of the ionised gas probed by the soft X-rays is in a ultra-ionised state, with IPs larger than $\sim 200$~eV. The signature left by such a ultra-ionised gas would lie predominantly in the soft X-ray energy range, which is already heavily absorbed by oxygen and carbon (both neutral and ionised). Resolving the absorption from an ultra-ionised gas is, therefore, beyond the spectral capabilities of current fast-response X-ray telescopes and will require a future era of X-ray telescopes such as {\it Xenia} \citep{pkd+10}.

Another possibility for detecting this ultra-ionised gas would be in the form of variability in emission lines as the gas cools on the order of years \citep[e.g.][]{prl00}. To date, analysis investigating such cooling has only been done in the case of GRB~990712 \citep{kgp06} where five epochs of observations of the host galaxy taken up to 6 years after the burst were used to search for variations in the emission line fluxes. No evidence of variability was found, from which the authors were able to put a limit on the density of the emitting (or ionised) region of $n\lesssim 6\times 10^3$~cm$^{-3}$. This remains consistent with our analysis.

\begin{figure}
\centering
\includegraphics[width=0.5\textwidth]{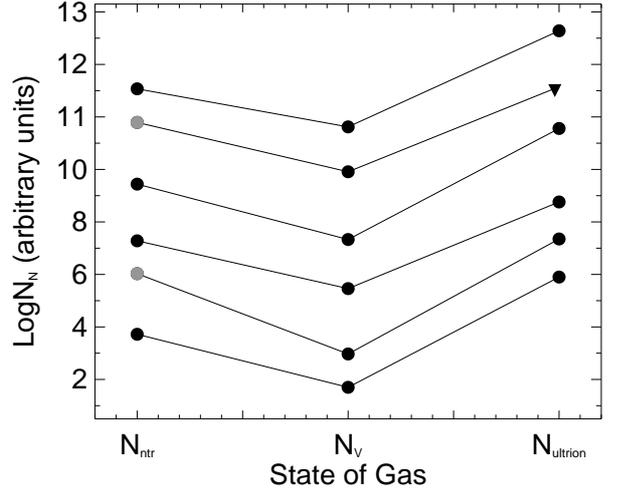}
\caption{Ionisation profile of gas within host galaxy of six GRBs as traced by nitrogen. Ionisation state of gas is divided into three broad components; neutral ($N_{\rm ntr}$), highly ionised ($N_V$), and ultra-ionised ($N_{\rm ultrion}$). Highly ionised nitrogen, N~{\sc v}, is measured directly from optical spectra, whereas the column density of neutral nitrogen is determined from either S~{\sc ii} (black circles) or Zn~{\sc ii} (grey circles), and the ultra-ionised column density of nitrogen gas is determined from X-ray measurements. The downward pointing triangle represents an upper-limit.} \label{fig:ionprof}
\end{figure}

\subsubsection{Gas Ionisation Profile}
\label{sec:disc}
A number of theoretical studies investigating the effect of a GRB on its surrounding environment show the intense radiation field produced by a GRB to be capable of destroying dust out to $\sim 10$~pc and ionising gas out to tens of parsecs \citep[e.g.][]{fkr01,pl02}. The large fraction of ionised gas along the line-of-sight to GRBs derived from our optical and X-ray absorption analysis may therefore not be that surprising. What may be of more surprise is the ionisation profile of the gas, which our analysis suggests has a well detected neutral component, a significantly smaller highly ionised component with IPs reaching up to $\sim 140$~eV, and then possibly a very large ultra-ionised component at IPs of $>200$~eV.

As an illustrative example we used the element nitrogen to visualise better the ionisation profile of the gas within GRB host galaxies along the line-of-sight, which is shown in Fig.~\ref{fig:ionprof}. Nitrogen is detected in a highly ionised form in six GRBs in our sample, and the neutral ($N_{\rm ntr}$) and ultra-ionised ($N_{\rm ultrion}$) fraction of nitrogen are derived from the column densities of neutral gas as traced by S~{\sc ii} (black circles) or Zn~{\sc ii} (grey circles), and from the gas column density inferred by soft X-ray absorption measurements, respectively. These column densities are then normalised to the cosmic abundances of nitrogren. An upper-limit on the ultra-ionised gas column density, $N_{\rm ultrion}$, in Fig~\ref{fig:ionprof} (top curve) is represented by a downwards pointing triangle.

The ionisation profile shown in Fig.~\ref{fig:ionprof} may arise from a distribution of gas with a density profile approximating a step-function, such that close to the GRB the density of the gas is significantly larger than far from the explosion sight, as suggested in section~\ref{sssec:seleff}. \citet{pl02} showed that a GRB can easily photoionise a dense column of gas ($>10^3$~cm$^{-3}$) out to a few parsecs. However, at lower densities further away from the GRB, the ionising cross-section of the gas drops, leaving the large fraction of the gas at these distances neutral. The ionisation potential bins used in Fig.~\ref{fig:ionprof} are, however, crude, and especially in the case of gas with IP $ > 200$~eV (i.e. $N_{\rm ultion}$), finer binning could produce a less pronounced dip in the ionisation profile at 20~eV--50~eV ($N_{\rm V}$) with an extended tail beyond 200~eV. In the context of the step-function-like density profile described above, a more extended, shallower density profile would by produced by a smoother density profile.

A more detailed investigation into the environmental conditions required to produce the observed ionisation profiles shown in Fig.~\ref{fig:ionprof} would require detailed modelling of the environment and radiation field produced by the GRB. Our ability to carry out such analysis at this time is, however, limited by the current lack of understanding of the conditions of the gas close to the GRB. Most of the lines measured in optical spectra trace material far from the GRB site, leaving us with little insight on the ionising flux of the radiation field, the structure and density of the environment, nor its temperature and metallicity. Greater constraint on the environmental properties within a few tens of parsecs to the GRB are, therefore, needed before an attempt to model the ionisation profile along the line-of-sight to the GRB can be made.

\section{Summary and Conclusions}
\label{sec:conc}
Using optical and X-ray spectra together with optical photometric afterglow data, we have combined the results from both spectroscopic and broadband SED analysis for a sample of 29~GRBs. This has allowed us to probe the properties of the host galaxy interstellar medium much deeper than would be possible with the analysis of optical or X-ray data alone.

From a sample of 18 GRBs we confirm the decreasing trend in the iron-to-zinc relative abundance as a function of $\log$~\Nzn\ previously observed \citep[e.g.][]{sav06}, and further find this correlation to be supported by independent dust extinction measurements from GRB afterglow SEDs. 

In comparing UV absorption lines of singly ionised metals, which are a tracer of neutral gas, to the soft X-ray absorption in the afterglows of 26 GRBs, which is representative of the total gas along the line-of-sight, we found the latter to be significantly larger. After taking careful consideration of systematic effects, we found the most likely interpretation for these observations to be that $\sim 90$\% of the host galaxy gas along the line-of-sight to the GRB is ionised.

Evidence for significant column densities of ionised gas within GRB host galaxies has already been presented, such as in \citet{pdr+08} and \citet{flv+08}. In both these papers, the velocity profiles of strong N~{\sc v} absorption in the host galaxies of a handful of GRBs were found to be consistent with arising from highly ionised circumburst gas photoionised by the GRB. However, from our analysis we found that only a negligible fraction ($\sim 0.01$~\%) of the X-ray absorbing gas is traced by the highly ionised metals measured in \citet{pdr+08} and \citet{flv+08}. This leads us to postulate that the majority of the ionised gas probed by our soft X-ray measurements is in an ultra-ionised state.

\citet{pdr+08} estimated that a cold gas ($T\approx 10^4$~K), with constant density profile would be mostly ionised by the GRB to N~{\sc v} out to a radius of $r\approx 10$~pc. The indication of our results that N~{\sc v} only traces a small fraction of the ionised gas within the GRB host galaxy thus suggests that most of the ionised gas lies inside the region where N~{\sc v} is detected, where the gas would experience a greater level of photoionisation.  The radius and density of the ionised cloud estimated by \citet{pdr+08} cannot account for the large column densities of ionised gas suggested by our X-ray measurements. However, both sets of observations may be satisfied by imposing a wind-density profile ($\ n\propto r^2$), such that a larger column of gas could be photoionised to very high states within a few parsecs of the GRB, while still maintaining a significant column of N~{\sc v}.

The recent surge in early time photometric and spectroscopic data available in the \swift\ era of GRB observations has provided a significant better handle on the conditions of the host galaxy material along the line-of-sight to a GRB, which shall be well complimented by  future observations with {\it XShooter} at VLT. Observations of variability in the fine-structure lines of singly ionised elements, such as Si~{\sc ii} and Fe~{\sc ii} are placing limits on the distances of the neutral absorbing gas \citep[e.g][]{vls+07,lvs+09}, and the detailed analysis of strong-ion absorption lines presented in \citet{flv+08} and \citet{pdr+08} provide some of the first attempts at investigating the GRB circumburst environment. 

From the broadband photometric and spectroscopic data presented in this paper we have been able study the relation between the neutral and ionised gas within GRB host galaxies, providing greater insight into the affect a GRB has on its surrounding environment, and on the importance of the numerous absorption components on our observations. The lack of optical absorption lines at lower IPs leads us to conclude that either the large fraction of the ionised gas along the line-of-sight to the GRB is in an ultra-ionised state, or that further consideration on our understanding of the absorption of the GRB optical and X-ray afterglow is needed. An ultra-ionised gas would produce a signature in the soft X-ray energy bands, where absorption from oxygen and carbon is already prominent. It will, however, require future, high sensitivity X-ray spectroscopic missions such as Xenia to verify the presence and abundance of such highly ionised gas.

\section*{ACKNOWLEDGEMENTS}
We thank the referee for the helpful comments. P. Schady acknowledges support through project SA 2001/2-1 of the DFG. S. Savaglio acknowledges support through project M.FE.A.Ext 00003 of the MPG. T. Kr{\"u}hler acknowledges support by the Deutsche Forschungsgemeinschaft (DFG) cluster of excellence ÕOrigin and Structure of the UniverseÕ. This research has made use of data obtained from the High Energy Astrophysics Science Archive Research Center (HEASARC), the UK Swift Science Data Centre at the University of Leicester and the Leicester Data base and Archive Service (LEDAS), provided by NASAs Goddard Space Flight Center and the Department of Physics and Astronomy, Leicester University, UK, respectively.

\end{document}